\documentclass[11pt]{article}
\usepackage[letterpaper,margin=1in]{geometry}
\usepackage[T1]{fontenc}
\usepackage{lmodern}
\usepackage{microtype}
\usepackage{graphicx}
\usepackage{booktabs}
\usepackage{amsmath}
\usepackage[numbers]{natbib}
\usepackage{upgreek,xspace}

\usepackage[colorlinks=true,linkcolor=blue,citecolor=blue,urlcolor=blue]{hyperref}
\usepackage{enumitem}
\setlist{nosep,leftmargin=*}
\makeatletter
\let\jnl@style=\rm
\def\ref@jnl#1{{\jnl@style#1}}

\def\aj{\ref@jnl{AJ}}                   
\def\actaa{\ref@jnl{Acta Astron.}}      
\def\araa{\ref@jnl{ARA\&A}}             
\def\apj{\ref@jnl{ApJ}}                 
\def\apjl{\ref@jnl{ApJ}}                
\def\apjs{\ref@jnl{ApJS}}               
\def\ao{\ref@jnl{Appl.~Opt.}}           
\def\apss{\ref@jnl{Ap\&SS}}             
\def\aap{\ref@jnl{A\&A}}                
\def\aapr{\ref@jnl{A\&A~Rev.}}          
\def\aaps{\ref@jnl{A\&AS}}              
\def\azh{\ref@jnl{AZh}}                 
\def\baas{\ref@jnl{BAAS}}               
\def\bac{\ref@jnl{Bull. astr. Inst. Czechosl.}}
\def\caa{\ref@jnl{Chinese Astron. Astrophys.}}
\def\cjaa{\ref@jnl{Chinese J. Astron. Astrophys.}}
\def\icarus{\ref@jnl{Icarus}}           
\def\jcap{\ref@jnl{J. Cosmology Astropart. Phys.}}
\def\jrasc{\ref@jnl{JRASC}}             
\def\memras{\ref@jnl{MmRAS}}            
\def\mnras{\ref@jnl{MNRAS}}             
\def\na{\ref@jnl{New A}}                
\def\nar{\ref@jnl{New A Rev.}}          
\def\pra{\ref@jnl{Phys.~Rev.~A}}        
\def\prb{\ref@jnl{Phys.~Rev.~B}}        
\def\prc{\ref@jnl{Phys.~Rev.~C}}        
\def\prd{\ref@jnl{Phys.~Rev.~D}}        
\def\pre{\ref@jnl{Phys.~Rev.~E}}        
\def\prl{\ref@jnl{Phys.~Rev.~Lett.}}    
\def\pasa{\ref@jnl{PASA}}               
\def\pasp{\ref@jnl{PASP}}               
\def\pasj{\ref@jnl{PASJ}}               
\def\qjras{\ref@jnl{QJRAS}}             
\def\rmxaa{\ref@jnl{Rev. Mexicana Astron. Astrofis.}}%
\def\rnaas{\ref@jnl{RNAAS}}             
\def\skytel{\ref@jnl{S\&T}}             
\def\solphys{\ref@jnl{Sol.~Phys.}}      
\def\sovast{\ref@jnl{Soviet~Ast.}}      
\def\ssr{\ref@jnl{Space~Sci.~Rev.}}     
\def\zap{\ref@jnl{ZAp}}                 
\def\nat{\ref@jnl{Nature}}              
\def\natas{\ref@jnl{Nat.~Ast.}}         
\def\iaucirc{\ref@jnl{IAU~Circ.}}       
\def\aplett{\ref@jnl{Astrophys.~Lett.}} 
\def\apspr{\ref@jnl{Astrophys.~Space~Phys.~Res.}}
\def\bain{\ref@jnl{Bull.~Astron.~Inst.~Netherlands}} 
\def\fcp{\ref@jnl{Fund.~Cosmic~Phys.}}  
\def\gca{\ref@jnl{Geochim.~Cosmochim.~Acta}}   
\def\grl{\ref@jnl{Geophys.~Res.~Lett.}} 
\def\jcp{\ref@jnl{J.~Chem.~Phys.}}      
\def\jgr{\ref@jnl{J.~Geophys.~Res.}}    
\def\jqsrt{\ref@jnl{J.~Quant.~Spec.~Radiat.~Transf.}}
\def\memsai{\ref@jnl{Mem.~Soc.~Astron.~Italiana}}
\def\nphysa{\ref@jnl{Nucl.~Phys.~A}}   
\def\physrep{\ref@jnl{Phys.~Rep.}}   
\def\physscr{\ref@jnl{Phys.~Scr}}   
\def\planss{\ref@jnl{Planet.~Space~Sci.}}   
\def\procspie{\ref@jnl{Proc.~SPIE}}   

\def\rprphys{\ref@jnl{Rep.~Prog.~Phys\@}}   
\def\ptp{\ref@jnl{Prog.~Th.~Phys.}}   
\def\natas{\ref@jnl{NatAs}}           
\def\amjm{\ref@jnl{AmJM}}             
\def\jatis{\ref@jnl{J.\ Astron.\ Tel.\ Instr.\ Syst.\@}} 

\makeatother

\title{\textbf{Beyond the Fast Rotators: A JWST Treasury Survey of Substellar Weather Across Rotation, Temperature and Viewing Geometry}}

\title{Beyond the Fast Rotators: A JWST Multi-Cycle
Treasury Survey of Substellar Weather}

\author{}
\date{}
\begin{document}
\maketitle
\begin{minipage}{\textwidth}
\centering

{\small \textbf{Natalia Oliveros-Gomez}$^{1}$, 
Elena Manjavacas$^{2,1}$,
Niall Whiteford$^{3}$,
Caprice Phillips$^{4}$,
Jared Bull$^{1}$,
Kielan Hoch$^{5}$,
Allison M. McCarthy$^{6}$,
Beatriz Campos Estrada$^{7}$,
Cian O'Toole$^{6}$,
Evert Nasedkin$^{6}$,
Stanimir Metchev$^{8}$,
Gabriel-Dominique Marleau$^{9,7}$,
Harshil Kothari$^{10}$, 
Antonia von Stauffenberg$^{7}$,
Michael K. Plummer$^{11}$,
Brianna I. Lacy$^{12}$,
Fuda Nguyen$^{13}$,
Xueqing Chen$^{14}$,
Fei Wang$^{15}$,
Genaro Su\'arez$^3$,
Ben J. Sutlieff$^{14}$
}

\vspace{5pt}

{\scriptsize
$^{1}$William H. Miller III Department of Physics and Astronomy,
Johns Hopkins University, Baltimore, MD 21218, USA\par
$^{2}$AURA for the European Space Agency (ESA), ESA Office, Space Telescope Science Institute, 3700 San Martin Drive\par
$^{3}$Department of Astrophysics, American Museum of Natural History,
New York, NY 10024\par
$^{4}$Department of Astronomy and Astrophysics, University of California, Santa Cruz, 1156 High Street, Santa Cruz, CA 95064\par
$^{5}$NASA Postdoctoral Fellow, Goddard Space Flight Center, 8800Greenbelt Rd, Greenbelt,MD 20771\par
$^{6}$School of Physics, Trinity College Dublin, The University of Dublin, Dublin 2, Ireland\par
$^{7}$ Max-Planck-Institut für Astronomie, Königstuhl 17, D-69117 Heidelberg, Germany \par
$^{8}$Department of Physics \& Astronomy, The University of Western Ontario, London, ON N6A 3K7, Canada\par
$^{9}$Universit\"at Duisburg--Essen, Lotharstra\ss{}e~1, 47057 Duisburg, Germany\par
$^{10}$Ritter Astrophysical Research Center, Department of Physics \& Astronomy, University of Toledo, 2801 W. Bancroft St., Toledo, OH 43606, USA\par
$^{11}$Department of Physics and Meteorology, United States Air Force Academy, 2354 Fairchild Drive, CO 80840, USA\par
\footnotesize
$^{12}$NASA Ames Research Center via Oak Ridge Associated Universities, 245 Pollack Rd., Moffett Field, CA 94035\par
$^{13}$Lunar \& Planetary Laboratory, University of Arizona, 1629 E. University Blvd.,
Tucson, AZ 85721\par
$^{14}$Institute for Astronomy, University of Edinburgh, Royal Observatory, Blackford Hill, Edinburgh, EH9 3HJ, UK\par
$^{15}$Department of Physics, Astronomy and Mathematics, University of Hertfordshire, AL10 9AB, Hatfield, UK\par
}

\vspace{4pt}

{\footnotesize Contact: \texttt{noliver8@jh.edu}\par}

\end{minipage}%

\begin{abstract}
\noindent
Atmospheric variability provides a unique window into the three-dimensional structure, dynamics, and evolution of brown dwarf and giant exoplanet atmospheres. Recent JWST time-resolved spectroscopic observations have revealed complex wavelength-dependent variability, offering unprecedented insights into cloud structures, thermal perturbations, and atmospheric chemistry. However, existing programs have focused primarily on individual benchmarks and specific scientific questions, resulting in heterogeneous spectral and temporal coverage and leaving important regions of the substellar population underexplored, particularly across a broad range of rotation periods, spectral types, and viewing geometries. A JWST Multi-Cycle Treasury program could systematically investigate substellar weather across a diverse, well-characterized population. By combining broad infrared spectroscopic coverage with observations spanning multiple rotations and strategically selected epochs, this program would constrain the contributions of different physical mechanisms while investigating how atmospheric structures depend on rotation, viewing geometry, and fundamental physical properties. Extended monitoring would also improve rotational-period determinations and reveal how atmospheric features evolve from rotation to rotation and over longer timescales. This coordinated effort would provide a consistently designed and uniformly processed dataset of time-resolved spectra and variability diagnostics, enabling population-level comparisons, testing predictions from three-dimensional circulation models, and establishing a framework for interpreting the atmospheric diversity and evolution of brown dwarfs and directly imaged giant exoplanets.
\end{abstract}

\newpage

\section{Introduction}

Atmospheric dynamics shape the observed properties of planetary atmospheres, as illustrated by the evolving cloud bands, storms, and vortices of Solar System gas giants \cite{showman2019atmospheric, rodriguez2025jwst}. Brown dwarfs provide an opportunity to investigate analogous processes beyond the Solar System, across a much broader range of temperatures, gravities, and rotational regimes. However, unlike Solar System planets, their atmospheres remain spatially unresolved, making time-resolved observations essential for probing their three-dimensional structure and evolution. Over the past decades, photometric and spectroscopic surveys have established atmospheric variability as a widespread phenomenon among brown dwarfs (e.g., \cite{koen2013extensive, radigan2014independent, buenzli2014brown, metchev2015weather, vos2020spitzer}). Ground-based monitoring, HST, and Spitzer have revealed diverse variability amplitudes, rotation periods, and evolving light-curve morphologies, while JWST now enables their characterization across a much broader infrared wavelength range, probing a wider range of atmospheric pressures.

Despite these advances, our understanding of the physical mechanisms driving substellar variability remains incomplete. Heterogeneous condensate clouds are commonly invoked to explain rotationally modulated brightness variations (\cite{Artigau2009, marley2010patchy, apai2013hst}), while thermal perturbations, atmospheric circulation, and thermochemical disequilibrium may also contribute (\cite{tremblin2015fingering, lee2023dynamically, mccarthy2024jwst, oliveros2025jwst}). Recent JWST observations have begun to reveal wavelength- and molecule-dependent variability that cloud heterogeneity alone cannot always explain, highlighting the complexity of the underlying atmospheric processes \cite{biller2013weather, mccarthy2024jwst, chen2024global, oliveros2025jwst, Miles2026, nasedkin2025jwst, Schrader2026}.

A comprehensive understanding of substellar weather requires moving beyond individual benchmark objects and limited observational configurations. The diversity of rotation periods, viewing geometries, atmospheric properties, and light-curve morphologies raises fundamental questions about how atmospheric structures form, evolve, and manifest in time-resolved observations. Addressing these questions requires a coordinated observational framework that samples a broader range of physical parameters, combines complementary atmospheric diagnostics, and characterizes variability over appropriate temporal baselines. A JWST Multi-Cycle Treasury program envisions a total of order 1,500 hours, sampling spectral type (L, T, Y), rotation (2-22~hr, \cite{tannock2021weather}), and viewing geometry (close to pole-on or equator objects), with repeated multi-rotation observations of selected targets across cycles to study the weather and climate of substellar objects.

\section{Scientific Motivation}
How do rotation, viewing geometry, and atmospheric evolution shape the diversity of substellar weather and climate, and how can JWST disentangle the physical mechanisms driving this variability across the brown dwarf population?

\subsection{Atmospheric dynamics, cloud coverage, temperature variations, and chemical disequilibrium}
\begin{figure}
    \centering
    \includegraphics[width=\linewidth]{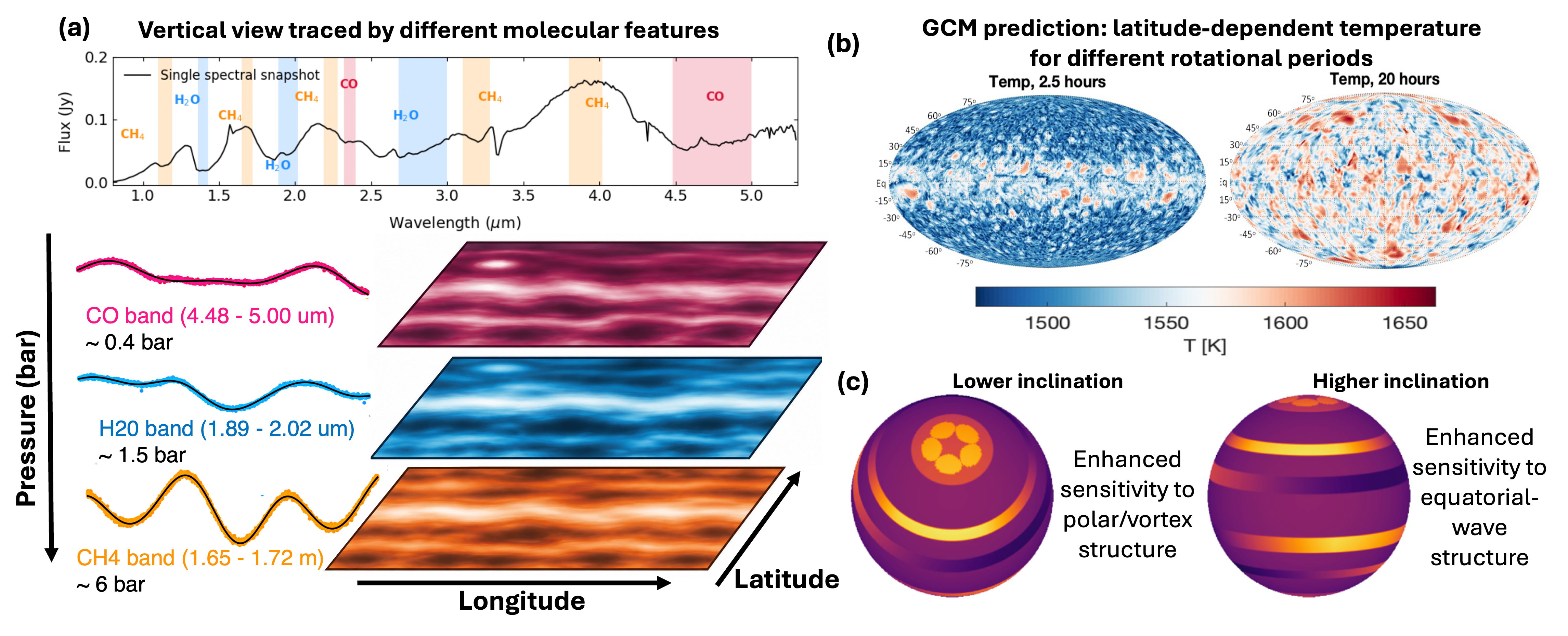}
    \caption{\small Connecting spectroscopic variability to three-dimensional atmospheric dynamics. (a, left) Single spectrum from the time series and three wavelength-dependent light curves of WISE 1049B, each from a spectrum snapshot, plus schematic atmospheric structures at different pressure levels traced by CO, H$_2$O, and CH$_4$  \cite{oliveros2025jwst}. (b, right top) Predicted temperature structures from atmospheric circulation models at different evolutionary times \cite{tan2021atmospheric-a}, and (c,  right bottom) schematic examples of polar vortices and equatorial cloud bands viewed at different inclinations \cite{fuda2024latitude}.}
    \label{fig:placeholder}
\end{figure}

Recent JWST time-resolved spectroscopy has revealed complex wavelength- and pressure-dependent variability that heterogeneous clouds alone cannot always explain. In WISE 1049A, CO and CH$_4$ show greater variability than H$_2$O, even at comparable atmospheric pressures (\cite{oliveros2025jwst}, Fig.~\ref{fig:placeholder}a). Similarly, harmonic analyses of SIMP 0136 identify distinct variability patterns across atmospheric depths and cloud- and molecule-sensitive spectral regions (\cite{mccarthy2024jwst, plummer2026mapping}). These findings suggest contributions from thermal and chemical processes in addition to cloud heterogeneity, highlighting the challenges of disentangling their relative roles.
Meanwhile, mid-infrared observations provide direct constraints on silicate clouds through their prominent absorption features at $\sim7-11~\mu$m, offering insight into cloud composition, particle properties, and vertical structure \citep{burningham2021cloud, luna2021empirically, Suarez2023}.

Complementary analyses using principal component analysis (PCA) have identified temperature variations and changes in cloud vertical structure as the dominant contributors to the observed spectroscopic variability of SIMP 0136 (\cite{akhmetshyn2025mapping, Schrader2026}). Moreover, brown dwarfs with similar fundamental properties can exhibit different spectra and atmospheric cloud characteristics, raising questions about how these processes depend on intrinsic atmospheric conditions and global circulation \citep{tan2021atmospheric-a, Suarez2023}. Combining JWST/NIRSpec molecular diagnostics with MIRI measurements of cloud-sensitive features across a diverse sample would allow us to investigate how cloud structure, thermal perturbations, and chemical processes jointly shape substellar variability, providing observational constraints for atmospheric retrievals and three-dimensional circulation models.

\subsection{Rotation, inclination, and effective temperature}

Rotation and viewing geometry are fundamental to interpreting substellar atmospheric variability. Observational studies find that brown dwarfs viewed closer to equator-on tend to show larger near-infrared variability amplitudes, redder colors, and stronger silicate absorption than those viewed closer to pole-on \cite{Vos2017, Suarez2023, fuda2024latitude, lam2026clouds}. Atmospheric circulation models predict distinct dynamical regimes, including zonal jets, atmospheric waves, and vortices, whose spatial distributions and observable variability depend on rotation and atmospheric conditions \cite{tan2021atmospheric-a, hammond2023shallow}. Recent models extending to Y dwarfs further explore the interplay between rotation, convection, and cloud radiative feedback in the coldest substellar atmospheres \cite{akin2026dynamics}. Together, these predictions motivate observations across a broader range of rotation periods, temperatures, and viewing geometries to test how atmospheric dynamics translate into observed light-curve morphologies. 

However, these trends come from relatively small, heterogeneous samples, making it difficult to disentangle viewing geometry from intrinsic atmospheric properties. Moreover, the relationship between rotation rate, atmospheric circulation, and observed light-curve morphology remains poorly constrained, particularly across different rotational regimes, and often shows strong degeneracies. Understanding these effects requires determining whether differences in variability amplitudes, wavelength-dependent behavior, and light-curve complexity arise from distinct atmospheric structures or from projection effects along our line of sight. A systematic JWST time-resolved spectroscopic survey spanning a broader range of rotation periods, and atmospheric conditions would enable direct comparisons of wavelength-dependent variability and morphology. Where available, independent inclination constraints would add another dimension for interpreting variability trends and testing the viewing-angle dependence predicted by atmospheric circulation models, without requiring inclination measurements for the entire sample.

\subsection{Atmospheric evolution over multiple timescales}

\begin{figure}[h!]
    \centering
    \includegraphics[width=\linewidth]{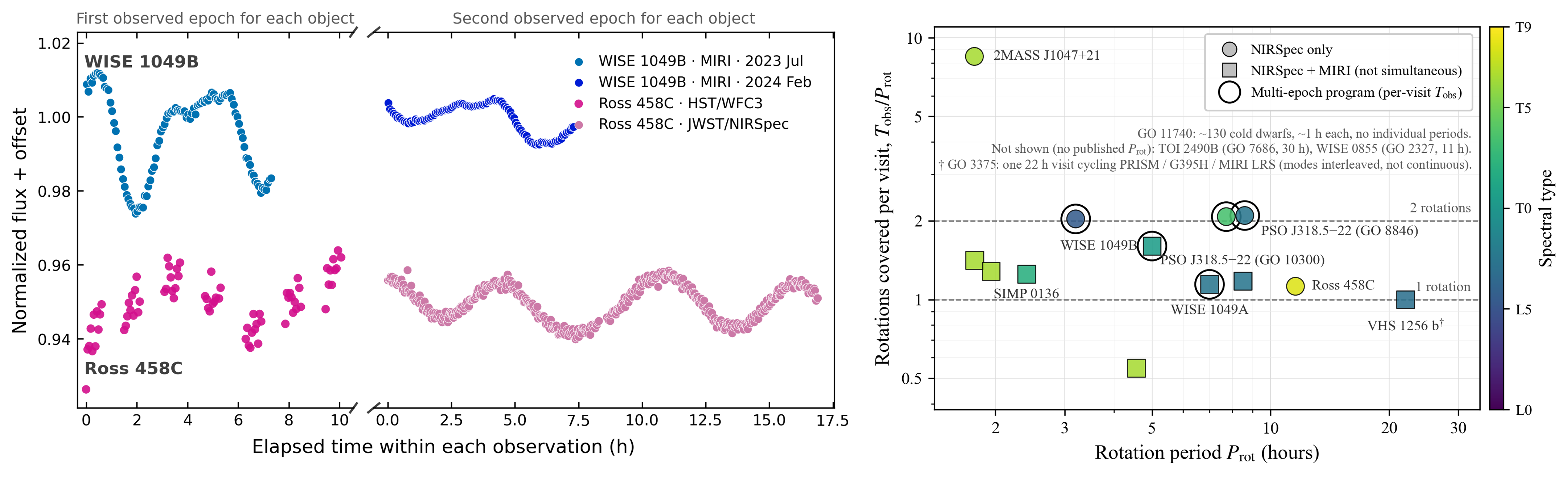}
    \caption{\small Diversity of atmospheric variability and rotational coverage of JWST observations of brown dwarfs. Left: Examples of light-curve evolution in WISE 1049B (5-12 $\mu$m) and Ross 458c (1.10-1.63 $\mu$m), illustrating changes in variability amplitude and morphology across observing epochs. \textit{Data taken from \cite{biller2024jwst, chen2025jwst, manjavacas2019cloud, manjavacas2026multi}.} Right: Rotational coverage of JWST time-resolved spectroscopic programs across Cycles 1–5, expressed as the ratio of the observing baseline per visit ($T_{\rm obs}$) to the rotational period ($P_{\rm rot}$). Marker colors indicate spectral type, while shapes distinguish NIRSpec-only observations (circles) from programs combining NIRSpec (PRISM, G395H, G235H) and MIRI (MRS and LRS, in squares). Open black outlines identify multi-epoch programs. Dashed horizontal lines indicate one and two rotational periods.}
    \label{fig:jwst_coverage}
\end{figure}

Multi-epoch observations reveal striking differences in substellar weather persistence. WISE 1049B exhibits evolving light-curve morphologies over seven months \cite{chen2025jwst}, whereas Ross 458c shows remarkably similar variability patterns across more than seven years (\cite{manjavacas2026multi}; Fig.~\ref{fig:jwst_coverage}). Multi-epoch HST monitoring of the planetary-mass companion VHS 1256 b also reveals substantial changes in amplitude and morphology, with variability reaching up to 38\% in the $J$-band \cite{zhou2022hst}. More recently, JWST/NIRISS and NIRSpec observations of SIMP 0136, separated by approximately 15 rotations, have revealed different wavelength-dependent variability patterns and atmospheric structures \cite{akhmetshyn2025mapping, plummer2026mapping, Schrader2026}.

These results demonstrate that atmospheric variability can evolve on timescales from a few rotations to years, although the physical mechanisms controlling such evolution remain poorly constrained, particularly in the coolest T/Y dwarfs. Continuous monitoring over multiple rotations is also needed to resolve periodicity ambiguities, as illustrated by revised rotation periods for Ross 458C \cite{manjavacas2026multi} and 2MASS J1721+33 \cite{o2026disentangling}. Together with archival HST and Spitzer observations, repeated JWST spectroscopy across rotations, months, and years would probe the evolution of molecular and cloud-sensitive features, testing models that predict atmospheric waves and slowly evolving structures on distinct timescales (\cite{fuda2024latitude, fuda2024polar}).

\subsection{Connecting to Exoplanets, Solar System weather and future facilities}
Brown dwarfs provide accessible laboratories for studying atmospheric processes shared with giant exoplanets and Solar System planets, such as vortices, cloud coverage, and temperature fluctuations  \cite{simon2016neptune, kaspi2023observational}, while the coldest T/Y dwarfs extend these comparisons toward Jupiter-like temperature regimes \cite{faherty2026, kuhnle2025water}. Beyond substellar science, a homogeneous JWST variability dataset would provide a benchmark for developing and validating atmospheric models, spectral decomposition methods, and time-series retrieval techniques. These methods will provide a valuable testing ground for future high-precision exoplanet spectroscopy, including observations with extremely large telescopes and future space missions, where reliable interpretations of atmospheric composition and potential biosignatures will require careful treatment of spectral degeneracies, atmospheric heterogeneity, and temporal variability.

\section{Why JWST, and why a Multi-Cycle Treasury program?}

JWST has transformed the study of substellar weather through high-precision, time-resolved spectroscopy spanning molecular and cloud-sensitive wavelengths. Programs approved across Cycles 1--5 have revealed complex wavelength-dependent variability in benchmark objects such as WISE 1049AB (GO 2965, (e.g., \cite{biller2024jwst, oliveros2025jwst, chen2025jwst})), VHS 1256 b (GO 3375, \cite{whiteford2026jwst}), SIMP 0136 (GO 3548, \cite{mccarthy2024jwst, Schrader2026}) extending to colder objects such as Ross 458c (GO 5226, \cite{manjavacas2026multi}) and WISE~0855 (GO 2327, \cite{Miles2026, kothari2026time}). Other programs have targeted objects with known magnetic activity, such as the radio-emitting T dwarf 2M~J1047+21 (GO 3496 and GO 6474), broadening the range of atmospheric phenomena accessible to JWST time-resolved spectroscopy. 

However, as Fig.~\ref{fig:jwst_coverage} right illustrates, detailed spectroscopic variability observations remain unevenly distributed across spectral types and rotational regimes. Early-L dwarfs are largely absent from the current JWST sample, limiting comparisons between warmer, cloud-dominated atmospheres and the evolving cloud structures at the L/T transition. At the other extreme, the coldest T/Y dwarfs remain poorly characterized in time-resolved spectroscopy, despite their importance for understanding cloud formation, atmospheric chemistry, and dynamics in colder, more planet-like conditions. The survey (GO 11740) could provide an important first step toward detecting variability in a larger sample of Y dwarfs, but its short ($\sim$1 hr) observations cannot generally characterize full rotational cycles or atmospheric evolution. These gaps, together with the heterogeneous wavelength and temporal coverage of existing programs, motivate a coordinated investigation across the substellar sequence.

Despite this substantial investment, existing JWST programs employ diverse observing modes, including NIRSpec/PRISM, G235H, G395H, and G395M, together with MIRI/LRS and MRS, providing complementary but heterogeneous spectral and temporal coverage. These differences complicate direct comparisons across the population and motivate a standardized observing strategy.




\section{Proposed Program and Estimated Observing Time}

We envision a stratified sample spanning the L, T, and Y spectral sequence, selected to explore the combined effects of atmospheric regime, rotation, and viewing geometry. Rather than maximizing the number of targets, the program would aim to provide meaningful coverage of fast, intermediate, and slow rotators, with a subset of independently constrained inclinations spanning approximately pole-on, intermediate, and equator-on orientations. Warmer early-L dwarfs would probe cloud-dominated atmospheres, while late-T and Y dwarfs would extend these comparisons toward colder, giant-planet-like conditions.

\textbf{Observing strategy and atmospheric characterization.} NIRSpec/PRISM time-resolved spectroscopy would form the core of the program, providing consistent molecular and continuum variability diagnostics across the sample. Complementary MIRI/LRS or MRS spectroscopy at each observing epoch would characterize mid-infrared atmospheric opacities, including silicate absorption in cloudy L dwarfs. A selected subset, particularly near the L/T transition, would additionally receive MIRI time-series spectroscopy to probe rotational changes in silicate absorption and cloud properties. Although NIRSpec and MIRI observe sequentially, their complementary wavelength coverage and sensitivity to different atmospheric pressures would provide stronger constraints on the physical mechanisms driving substellar variability.

\textbf{Multi-epoch atmospheric evolution.} Approximately half of the sample would be revisited at three epochs in total: initially, several months later, and more than one year after the first observation (following similar strategy of GO~8846). Each visit would aim to cover two rotations, allowing tracking of changes in variability amplitude, morphology, and wavelength-dependent behavior across timescales. Existing HST, Spitzer, and JWST observations would provide additional temporal benchmarks and guide target selection.

\textbf{Analysis framework and community legacy.} The program would deliver uniformly reduced time-resolved spectra, spectroscopic light curves, and standardized variability measurements. These datasets would support principal component analysis, time-series atmospheric retrievals, light-curve decomposition and classification, and comparisons with atmospheric circulation models (\cite{tan2021atmospheric-a, apai2017exploring, fuda2024latitude, Schrader2026}). The resulting public products would provide a consistent benchmark for testing atmospheric models across the substellar population.

\textbf{Program scale and the need for a Multi-Cycle Treasury allocation.} For an illustrative sample of 30 objects with a mean rotational period of 9~hr (spaning 2-20~hr, \cite{tannock2021weather}), two consecutive rotations of NIRSpec/PRISM spectroscopy would require approximately 540 hours for the initial observations. Two additional epochs for half the sample would add another 540 hours. Complementary MIRI/LRS or MRS spectroscopy at each epoch for all objects or one rotation of MIRI/LRS for selected L/T transition objects would bring the total to approximately 60-250~hr of mid-infrared observations, with durations depending on target brightness and instrumental configuration. Including these observations and operational overheads, the anticipated allocation would be on the order of 1500 hours under preliminary assumptions. The final time request would depend on the rotational-period distribution, sensitivity requirements, and observing feasibility. This coordinated combination of population coverage and repeated monitoring across multiple years would establish a legacy dataset that would be difficult to achieve through independent GO programs. 


\bibliographystyle{apj_short_prop.bst}
\bibliography{sample631}

@article{koen2013extensive,
  title={An extensive search for rapid optical variability in ultracool dwarfs},
  author={Koen, Chris},
  journal={\mnras},
  volume={428},
  number={4},
  pages={2824--2839},
  year={2013},
  publisher={Oxford University Press}
}

@article{o2026disentangling,
  title={Disentangling auroral-, cloud-, and magnetic spot-driven variability in three early L dwarfs with HST/WFC3},
  author={O’Toole, C and Vos, JM and Nasedkin, EN and Pineda, JS and Kao, MM and Zhou, Y and Schrader, M and McCarthy, AM},
  journal={\aap},
  volume={710},
  pages={A22},
  year={2026},
  publisher={EDP Sciences}
}

@article{kothari2026time,
  title={Time-resolved JWST Retrieval Analysis of the Coldest Brown Dwarf: Temperature and Chemical Variations in WISE 0855- 07},
  author={Kothari, Harshil and Morley, Caroline V and Miles, Brittany E and Rowland, Melanie J and Batalha, Natasha and Cushing, Michael C and Skemer, Andrew J and Mang, James and Lacy, Brianna and Vos, Johanna M and others},
  journal={\apj},
  volume={1008},
  number={1},
  pages={131},
  year={2026},
  publisher={The American Astronomical Society}
}

@article{luna2021empirically,
  title={Empirically determining substellar cloud compositions in the era of the James Webb Space Telescope},
  author={Luna, Jessica L and Morley, Caroline V},
  journal={\apj},
  volume={920},
  number={2},
  pages={146},
  year={2021},
  publisher={The American Astronomical Society}
}

@ARTICLE{Miles2026,
       author = {{Miles}, Brittany E. and {Mang}, James and {Morley}, Caroline V. and {Skemer}, Andrew J.~I. and {Rowland}, Melanie and {Lacy}, Brianna and {Suarez}, Genaro and {Visscher}, Channon and {Apai}, D{\'a}niel and {Bjoraker}, Gordon and et al.},
        title = "{Water Cloud and Chemical Modulations in the Coldest Brown Dwarf}",
      journal = {arXiv e-prints},
         year = 2026,
        month = sep,
          eid = {arXiv:2609.20664},
        pages = {arXiv:2609.20664},
          doi = {10.48550/arXiv.2609.20664},
archivePrefix = {arXiv},
       eprint = {2609.20664},
 primaryClass = {astro-ph.EP},
       adsurl = {https://ui.adsabs.harvard.edu/abs/2026arXiv260920664M}
}

@article{lam2026clouds,
  title={Clouds with a silicate lining: Using JWST spectra to probe atmospheric diversity in young AB Dor L dwarfs},
  author={Lam, MB and Vos, JM and Su{\'a}rez, G and Hsu, C-C and Bickle, TP and Faherty, J and Gagn{\'e}, J and Bardalez Gagliuffi, D and Biller, B and Burningham, B and others},
  journal={Astronomy \& Astrophysics},
  volume={709},
  pages={A56},
  year={2026},
  publisher={EDP Sciences}
}

@article{simon2016neptune,
  title={Neptune’s dynamic atmosphere from Kepler K2 observations: implications for brown dwarf light curve analyses},
  author={Simon, Amy A and Rowe, Jason F and Gaulme, Patrick and Hammel, Heidi B and Casewell, Sarah L and Fortney, Jonathan J and Gizis, John E and Lissauer, Jack J and Morales-Juberias, Raul and Orton, Glenn S and others},
  journal={The Astrophysical Journal},
  volume={817},
  number={2},
  pages={162},
  year={2016},
  publisher={The American Astronomical Society}
}

@article{kaspi2023observational,
  title={Observational evidence for cylindrically oriented zonal flows on Jupiter},
  author={Kaspi, Yohai and Galanti, Eli and Park, RS and Duer, K and Gavriel, N and Durante, D and Iess, L and Parisi, M and Buccino, DR and Guillot, T and others},
  journal={Nature Astronomy},
  volume={7},
  number={12},
  pages={1463--1472},
  year={2023},
  publisher={Nature Publishing Group UK, London}
}

@article{zhou2022hst,
  title={HST/WFC3 Complete Phase-resolved Spectroscopy of White-dwarf-brown-dwarf Binaries WD 0137 and EPIC 2122},
  author={Zhou, Yifan and Apai, D{\'a}niel and Tan, Xianyu and Lothringer, Joshua D and Lew, Ben WP and Casewell, Sarah L and Parmentier, Vivien and Marley, Mark S and Xu, Siyi and Mayorga, LC},
  journal={The Astronomical Journal},
  volume={163},
  number={1},
  pages={17},
  year={2022},
  publisher={The American Astronomical Society}
}

@article{plummer2026mapping,
  title={Mapping the Cloud-driven Atmospheric Dynamics and Chemistry of an Isolated Exoplanet Analog with Harmonic Signatures},
  author={Plummer, Michael K and Cocchini, Francis P and Kearns, Peter A and McCarthy, Allison M and Artigau, {\'E}tienne and Cowan, Nicolas B and Akhmetshyn, Roman and Vos, Johanna M and Nasedkin, Evert and Visscher, Channon and others},
  journal={The Astronomical Journal},
  volume={171},
  number={3},
  pages={195},
  year={2026},
  publisher={The American Astronomical Society}
}

@article{akin2026dynamics,
  title={Dynamics of Y dwarf atmospheres},
  author={Ak{\i}n, C and Lee, Elspeth KH and Gkouvelis, L and Heng, K},
  journal={Astronomy \& Astrophysics},
  volume={708},
  pages={A110},
  year={2026},
  publisher={EDP Sciences}
}

@article{hammond2023shallow,
  title={Shallow-water modelling of the atmospheric circulation regimes of brown dwarfs and their observable features},
  author={Hammond, Mark and Mayne, Nathan J and Seviour, William JM and Lewis, Neil T and Tan, Xianyu and Mitchell, Dann},
  journal={Monthly Notices of the Royal Astronomical Society},
  volume={525},
  number={1},
  pages={150--163},
  year={2023},
  publisher={Oxford University Press}
}

@article{akhmetshyn2025mapping,
  title={Mapping atmospheric features of the planetary-mass brown dwarf SIMP 0136 with JWST NIRISS},
  author={Akhmetshyn, Roman and Artigau, {\'E}tienne and Cowan, Nicolas B and Plummer, Michael K and Wang, Fei and Burningham, Ben and Benneke, Bj{\"o}rn and Doyon, Ren{\'e} and Jayawardhana, Ray and Lafreniere, David and others},
  journal={The Astrophysical Journal},
  volume={993},
  number={2},
  pages={237},
  year={2025},
  publisher={The American Astronomical Society}
}

@article{rodriguez2025jwst,
  title={JWST observations of exogenic species on Jupiter: HCN, H2O, and CO2},
  author={Rodr{\'\i}guez-Ovalle, Pablo and Fouchet, Thierry and Cavali{\'e}, Thibault and Lellouch, Emmanuel and Fletcher, Leigh N and Harkett, Jake and Hue, Vincent and Benmahi, Bilal and de Pater, Imke},
  journal={Astronomy \& Astrophysics},
  volume={696},
  pages={A173},
  year={2025},
  publisher={EDP Sciences}
}

@article{showman2019atmospheric,
  title={Atmospheric circulation of brown dwarfs and Jupiter-and Saturn-like planets: zonal jets, long-term variability, and QBO-type oscillations},
  author={Showman, Adam P and Tan, Xianyu and Zhang, Xi},
  journal={The Astrophysical Journal},
  volume={883},
  number={1},
  pages={4},
  year={2019},
  publisher={The American Astronomical Society}
}

@article{manjavacas2026multi,
  title={Multi-Epoch Stability in the Rotational Modulation of the Planetary-Mass Companion Ross 458C},
  author={Manjavacas, Elena and Bull, Jared and Miles-Paez, Paulo and Oliveros-Gomez, Natalia and Karalidi, Theodora and Lew, Ben WP and Zhang, Zhoujian and Zhou, Yifan and Apai, Daniel and Tan, Xianyu},
  journal={arXiv preprint arXiv:2609.22072},
  year={2026}
}

@ARTICLE{whiteford2026jwst,
       author = {{Whiteford}, Niall and {Faherty}, Jacqueline K. and {Burningham}, Ben and {Vos}, Johanna M. and {Petrus}, Simon and {Patapis}, Polychronis and {Biller}, Beth A. and {Skemer}, Andrew and {Hinkley}, Sasha and {Calamari}, Emily and {Su{\'a}rez}, Genaro and {Cruz}, Kelle L. and {Miles}, Brittany E. and {Carter}, Aarynn L. and {Martinez}, Francisco A. and {Rowland}, Melanie J. and {Absil}, Olivier and {Adams}, Arthur D. and {Balmer}, William O. and {Boccaletti}, Anthony and {Bonavita}, Mariangela and {Bonnefoy}, Micka{\"e}l and {Booth}, Mark and {Bowler}, Brendan P. and {Briesemeister}, Zackery W. and {Bryan}, Marta L. and {Calissendorff}, Per and {Cantalloube}, Faustine and {Charnay}, Benjamin and {Chauvin}, Ga{\"e}l and {Chen}, Christine H. and {Choquet}, Elodie and {Christiaens}, Valentin and {Cugno}, Gabriele and {Currie}, Thayne and {Danielski}, Camilla and {De Furio}, Matthew and {Dupuy}, Trent J. and {Factor}, Samuel M. and {Fitzgerald}, Michael P. and {Fortney}, Jonathan J. and {Franson}, Kyle and {Girard}, Julien H. and {Gonzales}, Eileen C. and {Grady}, Carol A. and {Henning}, Thomas and {Hines}, Dean C. and {Hood}, Callie E. and {Hoch}, Kielan K.~W. and {Howe}, Alex R. and {Janson}, Markus and {Kalas}, Paul and {Kammerer}, Jens and {Kennedy}, Grant M. and {Kervella}, Pierre and {Kim}, Minjae and {Kitzmann}, Daniel and {Kraus}, Adam L. and {Kuzuhara}, Masayuki and {Lagage}, Pierre-Olivier and {Lagrange}, Anne-Marie and {Lawson}, Kellen and {Lazzoni}, Cecilia and {Leisenring}, Jarron M. and {Lew}, Ben W.~P. and {Liu}, Michael C. and {Liu}, Pengyu and {Llop-Sayson}, Jorge and {Lloyd}, James P. and {Lueber}, Anna and {Macintosh}, Bruce and {M{\^a}lin}, Mathilde and {Manjavacas}, Elena and {Marino}, Sebasti{\'a}n and {Marley}, Mark S. and {Marois}, Christian and {Martinez}, Raquel A. and {Matthews}, Elisabeth C. and {Matthews}, Brenda C. and {Mawet}, Dimitri and {Mazoyer}, Johan and {McElwain}, Michael W. and {Metchev}, Stanimir and {Meyer}, Michael R. and {Millar-Blanchaer}, Maxwell A. and {Molli{\`e}re}, Paul and {Moran}, Sarah E. and {Morley}, Caroline V. and {Mukherjee}, Sagnick and {Palma-Bifani}, Paulina and {Pantin}, Eric and {Perrin}, Marshall D. and {Pueyo}, Laurent and {Quanz}, Sascha P. and {Quirrenbach}, Andreas and {Ray}, Shrishmoy and {Rebollido}, Isabel and {Adams Redai}, Jea and {Ren}, Bin B. and {Rickman}, Emily and {Sallum}, Steph and {Samland}, Matthias and {Sargent}, Benjamin and {Schlieder}, Joshua E. and {Stapelfeldt}, Karl R. and {Stone}, Jordan M. and {Sutlieff}, Ben J. and {Tamura}, Motohide and {Tan}, Xianyu and {Theissen}, Christopher A. and {Tremblin}, Pascal and {Uyama}, Taichi and {Vasist}, Malavika and {Vigan}, Arthur and {Wagner}, Kevin and {Wang}, Jason J. and {Ward-Duong}, Kimberly and {Wolff}, Schuyler G. and {Worthen}, Kadin and {Wyatt}, Mark C. and {Ygouf}, Marie and {Zurlo}, Alice and {Zhang}, Xi and {Zhang}, Keming and {Zhang}, Zhoujian and {Zhou}, Yifan},
        title = "{The JWST Early Release Science Program for Direct Observations of Exoplanetary Systems VIII: patchy forsterite and enstatite clouds in the atmosphere of VHS 1256 b, retrieval lessons learned and outlook to the future}",
      journal = {arXiv e-prints},
         year = 2026,
        month = aug,
          eid = {arXiv:2608.06583},
        pages = {arXiv:2608.06583},
          doi = {10.48550/arXiv.2608.06583},
archivePrefix = {arXiv},
       eprint = {2608.06583},
 primaryClass = {astro-ph.EP},
       adsurl = {https://ui.adsabs.harvard.edu/abs/2026arXiv260806583W}
}

@article{marley2010patchy,
  title={A patchy cloud model for the L to T dwarf transition},
  author={Marley, Mark S and Saumon, Didier and Goldblatt, Colin},
  journal={\apjl},
  volume={723},
  number={1},
  pages={L117--L121},
  year={2010},
  publisher={The American Astronomical Society}
}

@article{kuhnle2025water,
  title={Water depletion and 15NH3 in the atmosphere of the coldest brown dwarf observed with JWST/MIRI},
  author={K{\"u}hnle, Helena and Patapis, Polychronis and Molli{\`e}re, Paul and Tremblin, Pascal and Matthews, Elisabeth and Glauser, Adrian Michael and Whiteford, Niall and Vasist, Malavika and Absil, Olivier and Barrado, David and others},
  journal={\aap},
  volume={695},
  pages={A224},
  year={2025},
  publisher={EDP Sciences}
}

@article{oliveros2025jwst,
  title={The JWST Weather Report from the Nearest Brown Dwarfs. III. Heterogeneous Clouds and Thermochemical Instabilities as Possible Drivers of WISE 1049AB’s Spectroscopic Variability},
  author={Oliveros-Gomez, Natalia and Manjavacas, Elena and Karalidi, Theodora and Phillippe, Myrla and Campos Estrada, Beatriz and Biller, Beth and Vos, Johanna M and Faherty, Jacqueline and Chen, Xueqing and Dupuy, Trent J and others},
  journal={\apj},
  volume={997},
  number={2},
  pages={136},
  year={2026},
  publisher={The American Astronomical Society}
}

@article{nasedkin2025jwst,
  title={The JWST weather report: Retrieving temperature variations, auroral heating, and static cloud coverage on SIMP-0136},
  author={Nasedkin, Evert and Schrader, Merle and Vos, Johanna M and Biller, Beth and Burningham, Ben and Cowan, Nicolas B and Faherty, JK and Gonzales, Eileen and Lam, Madeline B and McCarthy, Allison M and others},
  journal={\aap},
  volume={702},
  pages={A1},
  year={2025},
  publisher={EDP Sciences}
}

@article{biller2013weather,
  title={WEATHER ON THE NEAREST BROWN DWARFS: RESOLVED SIMULTANEOUS MULTI-WAVELENGTH VARIABILITY MONITORING OF WISE J104915. 57- 531906.1 AB},
  author={Biller, Beth A and Crossfield, Ian JM and Mancini, Luigi and Ciceri, Simona and Southworth, John and Kopytova, Taisiya G and Bonnefoy, Micka{\"e}l and Deacon, Niall R and Schlieder, Joshua E and Buenzli, Esther and others},
  journal={\apjl},
  volume={778},
  number={1},
  pages={L10},
  year={2013},
  publisher={IOP Publishing}
}

@article{chen2025jwst,
  title={The JWST weather report from the nearest brown dwarfs II: consistent variability mechanisms over 7 months revealed by 1--14 $\mu$m NIRSpec+ MIRI monitoring of WISE 1049AB},
  author={Chen, Xueqing and Biller, Beth A and Tan, Xianyu and Vos, Johanna M and Zhou, Yifan and Su{\'a}rez, Genaro and McCarthy, Allison M and Morley, Caroline V and Whiteford, Niall and Dupuy, Trent J and others},
  journal={\mnras},
  volume={539},
  number={4},
  pages={3758--3777},
  year={2025},
  publisher={Oxford University Press}
}

@article{chen2024global,
  title={Global weather map reveals persistent top-of-atmosphere features on the nearest brown dwarfs},
  author={Chen, Xueqing and Biller, Beth A and Vos, Johanna M and Crossfield, Ian JM and Mace, Gregory N and Hood, Callie E and Tan, Xianyu and Allers, Katelyn N and Martin, Emily C and Bubb, Emma and others},
  journal={\mnras},
  volume={533},
  number={3},
  pages={3114--3143},
  year={2024},
  publisher={Oxford University Press}
}

@article{fuda2024polar,
  title={The polar vortex hypothesis: Evolving, spectrally distinct polar regions explain short-and long-term light-curve evolution and color--inclination trends in brown dwarfs and giant exoplanets},
  author={Fuda, Nguyen and Apai, D{\'a}niel},
  journal={The Astrophysical Journal Letters},
  volume={975},
  number={2},
  pages={L32},
  year={2024},
  publisher={The American Astronomical Society}
}

@article{fuda2024latitude,
  title={Latitude-dependent Atmospheric Waves and Long-period Modulations in Luhman 16 B from the Longest Light Curve of an Extrasolar World},
  author={Fuda, Nguyen and Apai, D{\'a}niel and Nardiello, Domenico and Tan, Xianyu and Karalidi, Theodora and Bedin, Luigi Rolly},
  journal={\apj},
  volume={965},
  number={2},
  pages={182},
  year={2024},
  publisher={IOP Publishing}
}

@article{biller2024jwst,
  title={The JWST weather report from the nearest brown dwarfs I: multiperiod JWST NIRSpec+ MIRI monitoring of the benchmark binary brown dwarf WISE 1049AB},
  author={Biller, Beth A and Vos, Johanna M and Zhou, Yifan and McCarthy, Allison M and Tan, Xianyu and Crossfield, Ian JM and Whiteford, Niall and Suarez, Genaro and Faherty, Jacqueline and Manjavacas, Elena and others},
  journal={\mnras},
  volume={532},
  number={2},
  pages={2207--2233},
  year={2024},
  publisher={Oxford University Press}
}

@ARTICLE{Artigau2009,
       author = {{Artigau}, {\'E}tienne and {Bouchard}, Sandie and {Doyon}, Ren{\'e} and {Lafreni{\`e}re}, David},
        title = "{Photometric Variability of the T2.5 Brown Dwarf SIMP J013656.5+093347: Evidence for Evolving Weather Patterns}",
      journal = {\apj},
         year = 2009,
        month = aug,
       volume = {701},
       number = {2},
        pages = {1534-1539},
          doi = {10.1088/0004-637X/701/2/1534},
archivePrefix = {arXiv},
       eprint = {0906.3514},
 primaryClass = {astro-ph.SR},
       adsurl = {https://ui.adsabs.harvard.edu/abs/2009ApJ...701.1534A}
}

@ARTICLE{Suarez2023,
       author = {{Su{\'a}rez}, Genaro and {Vos}, Johanna M. and {Metchev}, Stanimir and {Faherty}, Jacqueline K. and {Cruz}, Kelle},
        title = "{Ultracool Dwarfs Observed with the Spitzer Infrared Spectrograph: Equatorial Latitudes in L Dwarf Atmospheres Are Cloudier}",
      journal = {\apjl},
         year = 2023,
        month = sep,
       volume = {954},
       number = {1},
          eid = {L6},
        pages = {L6},
          doi = {10.3847/2041-8213/acec4b},
archivePrefix = {arXiv},
       eprint = {2308.02093},
 primaryClass = {astro-ph.EP},
       adsurl = {https://ui.adsabs.harvard.edu/abs/2023ApJ...954L...6S}
}

@ARTICLE{Vos2017,
       author = {{Vos}, Johanna M. and {Allers}, Katelyn N. and {Biller}, Beth A.},
        title = "{The Viewing Geometry of Brown Dwarfs Influences Their Observed Colors and Variability Amplitudes}",
      journal = {\apj},
         year = 2017,
        month = jun,
       volume = {842},
       number = {2},
          eid = {78},
        pages = {78},
          doi = {10.3847/1538-4357/aa73cf},
archivePrefix = {arXiv},
       eprint = {1705.06045},
 primaryClass = {astro-ph.SR},
       adsurl = {https://ui.adsabs.harvard.edu/abs/2017ApJ...842...78V}
}

@article{tremblin2015fingering,
  title={Fingering convection and cloudless models for cool brown dwarf atmospheres},
  author={Tremblin, Pascal and Amundsen, David S and Mourier, P and Baraffe, I and Chabrier, G and Drummond, B and Homeier, D and Venot, Olivia},
  journal={\apjl},
  volume={804},
  number={1},
  pages={L17},
  year={2015},
  publisher={IOP Publishing}
}

@article{vos2020spitzer,
  title={Spitzer variability properties of low-gravity L dwarfs},
  author={Vos, Johanna M and Biller, Beth A and Allers, Katelyn N and Faherty, Jacqueline K and Liu, Michael C and Metchev, Stanimir and Eriksson, Simon and Manjavacas, Elena and Dupuy, Trent J and Janson, Markus and others},
  journal={\aj},
  volume={160},
  number={1},
  pages={38},
  year={2020},
  publisher={IOP Publishing}
}

@article{apai2013hst,
  title={HST spectral mapping of L/T transition brown dwarfs reveals cloud thickness variations},
  author={Apai, D{\'a}niel and Radigan, Jacqueline and Buenzli, Esther and Burrows, Adam and Reid, Iain Neill and Jayawardhana, Ray},
  journal={\apj},
  volume={768},
  number={2},
  pages={121},
  year={2013},
  publisher={IOP Publishing}
}

@ARTICLE{apai2017exploring,
       author = {{Apai}, Daniel and {Cowan}, Nicolas and {Kopparapu}, Ravikumar and {Kasper}, Markus and {Hu}, Renyu and {Morley}, Caroline and {Fujii}, Yuka and {Kane}, Stephen and {Maley}, Mark and {del Genio}, Anthony and {Karalidi}, Theodora and {Komacek}, Thaddeus and {Mamajek}, Eric and {Mandell}, Avi and {Domagal-Goldman}, Shawn and {Barman}, Travis and {Boss}, Alan and {Breckinridge}, James and {Crossfield}, Ian and {Danchi}, William and {Ford}, Eric and {Iro}, Nicolas and {Kasting}, James and {Lowrance}, Patrick and {Madhusudhan}, Nikku and {McElwain}, Michael and {Moore}, William and {Pascucci}, Ilaria and {Plavchan}, Peter and {Roberge}, Aki and {Schneider}, Glenn and {Showman}, Adam and {Turnbull}, Margaret},
        title = "{Exploring Other Worlds: Science Questions for Future Direct Imaging Missions (EXOPAG SAG15 Report)}",
      journal = {arXiv e-prints},
         year = 2017,
        month = aug,
          eid = {arXiv:1708.02821},
        pages = {arXiv:1708.02821},
archivePrefix = {arXiv},
       eprint = {1708.02821},
 primaryClass = {astro-ph.EP},
       adsurl = {https://ui.adsabs.harvard.edu/abs/2017arXiv170802821A}
}

@article{tannock2021weather,
  title={Weather on Other Worlds. V. The Three Most Rapidly Rotating Ultra-cool Dwarfs},
  author={Tannock, Megan E and Metchev, Stanimir and Heinze, Aren and Miles-P{\'a}ez, Paulo A and Gagn{\'e}, Jonathan and Burgasser, Adam and Marley, Mark S and Apai, D{\'a}niel and Su{\'a}rez, Genaro and Plavchan, Peter},
  journal={\aj},
  volume={161},
  number={5},
  pages={224},
  year={2021},
  publisher={IOP Publishing}
}

@article{radigan2014independent,
  title={An independent analysis of the brown dwarf atmosphere monitoring (BAM) data: large-amplitude variability is rare outside the L/T transition},
  author={Radigan, Jacqueline},
  journal={\apj},
  volume={797},
  number={2},
  pages={120},
  year={2014},
  publisher={IOP Publishing}
}

@article{manjavacas2019cloud,
  title={Cloud Atlas: Hubble Space Telescope Near-infrared Spectral Library of Brown Dwarfs, Planetary-mass Companions, and Hot Jupiters},
  author={Manjavacas, Elena and Apai, D{\'a}niel and Zhou, Yifan and Lew, Ben WP and Schneider, Glenn and Metchev, Stan and Miles-P{\'a}ez, Paulo A and Radigan, Jacqueline and Marley, Mark S and Cowan, Nicolas and others},
  journal={\aj},
  volume={157},
  number={3},
  pages={101},
  year={2019},
  publisher={IOP Publishing}
}

@ARTICLE{mccarthy2024jwst,
       author = {{McCarthy}, Allison M. and {Vos}, Johanna M. and {Muirhead}, Philip S. and {Biller}, Beth A. and {Morley}, Caroline V. and {Faherty}, Jacqueline and {Burningham}, Ben and {Calamari}, Emily and {Cowan}, Nicolas B. and {Cruz}, Kelle L. and {Gonzales}, Eileen and {Limbach}, Mary Anne and {Liu}, Pengyu and {Nasedkin}, Evert and {Su{\'a}rez}, Genaro and {Tan}, Xianyu and {O'Toole}, Cian and {Visscher}, Channon and {Whiteford}, Niall and {Zhou}, Yifan},
        title = "{The JWST Weather Report from the Isolated Exoplanet Analog SIMP 0136+0933: Pressure-dependent Variability Driven by Multiple Mechanisms}",
      journal = {\apjl},
         year = 2025,
        month = mar,
       volume = {981},
       number = {2},
          eid = {L22},
        pages = {L22},
          doi = {10.3847/2041-8213/ad9eaf},
archivePrefix = {arXiv},
       eprint = {2411.16577},
 primaryClass = {astro-ph.EP},
       adsurl = {https://ui.adsabs.harvard.edu/abs/2025ApJ...981L..22M}
}

@article{burningham2021cloud,
  title={Cloud busting: enstatite and quartz clouds in the atmosphere of 2M2224-0158},
  author={Burningham, Ben and Faherty, Jacqueline K and Gonzales, Eileen C and Marley, Mark S and Visscher, Channon and Lupu, Roxana and Gaarn, Josefine and Fabienne Bieger, Michelle and Freedman, Richard and Saumon, Didier},
  journal={\mnras},
  volume={506},
  number={2},
  pages={1944--1961},
  year={2021},
  publisher={Oxford University Press}
}

@article{metchev2015weather,
  title={Weather on other worlds. II. Survey results: Spots are ubiquitous on L and T dwarfs},
  author={Metchev, Stanimir A and Heinze, Aren and Apai, D{\'a}niel and Flateau, Davin and Radigan, Jacqueline and Burgasser, Adam and Marley, Mark S and Artigau, {\'E}tienne and Plavchan, Peter and Goldman, Bertrand},
  journal={\apj},
  volume={799},
  number={2},
  pages={154},
  year={2015},
  publisher={IOP Publishing}
}

@article{lee2023dynamically,
  title={Dynamically coupled kinetic chemistry in brown dwarf atmospheres--I. Performing global scale kinetic modelling},
  author={Lee, Elspeth KH and Tan, Xianyu and Tsai, Shang-Min},
  journal={\mnras},
  volume={523},
  number={3},
  pages={4477--4491},
  year={2023},
  publisher={Oxford University Press}
}

@article{tan2021atmospheric-a,
  title={Atmospheric circulation of brown dwarfs and directly imaged exoplanets driven by cloud radiative feedback: effects of rotation},
  author={Tan, Xianyu and Showman, Adam P},
  journal={\mnras},
  volume={502},
  number={1},
  pages={678--699},
  year={2021},
  publisher={Oxford University Press}
}

@article{buenzli2014brown,
  title={Brown dwarf photospheres are patchy: a hubble space telescope near-infrared spectroscopic survey finds frequent low-level variability},
  author={Buenzli, Esther and Apai, D{\'a}niel and Radigan, Jacqueline and Reid, I Neill and Flateau, Davin},
  journal={\apj},
  volume={782},
  number={2},
  pages={77},
  year={2014},
  publisher={IOP Publishing}
}

@ARTICLE{Faherty2026,
       author = {{Faherty}, Jacqueline K. and {Vos}, Johanna M.},
        title = "{Observed Characteristics of Brown Dwarf Atmospheres}",
      journal = {\araa},
         year = 2026,
        month = aug,
       volume = {64},
       number = {1},
        pages = {409-447},
          doi = {10.1146/annurev-astro-052722-103322},
       adsurl = {https://ui.adsabs.harvard.edu/abs/2026ARA&A..64..409F}
}

@ARTICLE{Schrader2026,
       author = {{Schrader}, Merle A. and {Vos}, Johanna M. and {Nasedkin}, Evert and {Kestell}, Jennifer and {Cowan}, Nicolas B. and {Akhmetshyn}, Roman and {Beiler}, Samuel and {Biller}, Beth A. and {Burningham}, Ben and {Faherty}, Jacqueline and {Gonzales}, Eileen C. and {McCarthy}, Allison M. and {Morley}, Caroline V. and {O'Donovan}, Barry and {O'Toole}, Cian and {Su{\'a}rez}, Genaro and {Tan}, Xianyu and {Visscher}, Channon and {Whiteford}, Niall and {Zhou}, Yifan},
        title = "{The JWST weather report: Unravelling the atmospheric variability of isolated worlds using principal component analysis}",
      journal = {\aap},
         year = 2026,
        month = sep,
       volume = {713},
          eid = {A192},
        pages = {A192},
          doi = {10.1051/0004-6361/202660109},
archivePrefix = {arXiv},
       eprint = {2607.26182},
 primaryClass = {astro-ph.EP},
       adsurl = {https://ui.adsabs.harvard.edu/abs/2026A&A...713A.192S}
}

\end{document}